\documentclass[onecolumn]{aastex}
\usepackage{graphicx}
\usepackage{color}
\usepackage{hyperref}
\usepackage{cite}
\usepackage{array}
\usepackage{tabularx}

\shorttitle{}

\shortauthors{Zhu et al.}

\def\degree{${}^{\circ}$}

\newcommand{\stereo}{\textit{STEREO}}
\newcommand{\stereoa}{\textit{STEREO A}}
\newcommand{\stereob}{\textit{STEREO B}}
\newcommand{\soho}{\textit{SOHO}}
\newcommand{\ace}{\textit{ACE}}
\newcommand{\sdo}{\textit{SDO}}
\newcommand{\wind}{\textit{Wind}}
\newcommand{\goes}{\textit{GOES}}
\newcommand{\gong}{\textit{GONG}}

\begin{document}

\title{Investigation of energetic particle release using multi-point imaging and in situ observations}

\author{Bei Zhu\altaffilmark{1,2}, Ying D. Liu\altaffilmark{1,2}, Ryun-Young Kwon\altaffilmark{3}, Rui Wang\altaffilmark{1}}

\altaffiltext{1}{State Key Laboratory of Space Weather, National Space
Science Center, Chinese Academy of Sciences, Beijing 100190, China;
liuxying@swl.ac.cn}

\altaffiltext{2}{University of Chinese Academy of Sciences, No.19A
Yuquan Road, Beijing 100049, China}

\altaffiltext{3}{College of Science, George Mason University, 4400 University Drive, Fairfax, VA 22030, USA}

\begin{abstract}
The solar eruption on 2012 January 27 resulted in a wide-spread solar energetic particle (SEP) event observed by \stereoa{} and the near-Earth spacecraft (separated by 108\degree).
The event was accompanied by an X-class flare, extreme-ultraviolet (EUV) wave and fast coronal mass ejection (CME).
We investigate the particle release by comparing the release times of particles at the spacecraft and the times when magnetic connectivity between the source and the spacecraft was established.
The EUV wave propagating to the magnetic footpoint of the spacecraft in the lower corona and the shock expanding to the open field line connecting the spacecraft in the upper corona are thought to be responsible for the particle release.
We track the evolution of the EUV wave and model the propagation of the shock using EUV and white-light observations.
No obvious evidence indicates that the EUV wave reached the magnetic footpoint of either \stereoa{} or L1-observers.
Our shock modeling shows that the release time of the particles observed at L1 was consistent with the time when the shock first established contact with the magnetic field line connecting L1-observers.
The release of the particles observed by \stereoa{} was delayed relative to the time when the shock was initially connected to \stereoa{} via the magnetic field line.
We suggest that the particle acceleration efficiency of the portion of the shock connected to the spacecraft determines the release of energetic particles at the spacecraft.

\end{abstract}

\section{Introduction}

A straightforward interpretation for a gradual solar energetic particle (SEP) event observed simultaneously by spacecraft at different locations in the heliosphere is the wide extent of a coronal mass ejection (CME)-driven shock in the corona or interplanetary (IP) medium \citep[e.g.,][]{Cliver95, Heras95, Reames99, Reames13}.
The particles are accelerated at the shock and then injected onto the magnetic field lines connecting the observers in the heliosphere \citep[e.g.,][]{Cliver04, Zank07,Kozarev15}.
Cross-field diffusion processes have also been suggested to account for the particle transport over a wide longitudinal range from a narrow particle source region \citep[e.g.,][]{Wang12,Lario17_Aug14}.
Studies based on the joint observations from the \emph{Solar-TErrestrial RElations Observatory} \citep[\stereo;][] {Kaiser08} and near-Earth spacecraft such as the \emph{Advanced Composition Explorer} \citep[\ace;][] {Stone98} and the \emph{Solar and Heliospheric Observatory} \citep[\soho;][] {Domingo95}, have been carried out for the longitudinal distribution and the release mechanism of energetic particles \citep[e.g.,][]{Rouillard12,Prise14, Richardson14, Gomez-Herrero15,Kouloumvakos15,Kozarev16,Rouillard16,Zhu16, Lario13, Lario14, Lario16, Lario17_Aug14, Kwon17}.
In addition, the remote-sensing observations from the \emph{Atmospheric Imaging Assembly} \citep[AIA;][] {Lemen12} on board the \emph{Solar Dynamics Observatory} \citep[\sdo;][] {Pesnell12} with high cadence and from the \emph{Sun-Earth Connection Coronal and Heliospheric Investigation} \citep[SECCHI;][] {Howard08} instruments on board \stereo{} offer an unprecedented opportunity to investigate the EUV wave and shock evolution and the connectivity between the observers and the coronal disturbances.

The properties of the CME-driven shock and accompanied physical phenomena are used to study the mechanisms of particle acceleration and release and hence explain the wide distribution of SEP events.
EUV waves, which propagate in the solar corona and can be detected in EUV images, are considered as one of the candidates to be responsible for the particle release in the SEP events.
\citet{Rouillard12} examine the SEP event on 2011 March 21 and conclude that the arrival times of the EUV wave at the magnetic footpoints of the spacecraft coincide with the particle release times.
\citet{Park13} suggest that the EUV wave traces the release site of SEPs accelerated by the CME-driven shock in a study of 12 SEP events.
Statistical work by \citet{Miteva14} with 179 SEP events show that a large majority of SEP events are associated with EUV waves.
They get a connectivity between the extrapolated arrival times of the EUV wave at the magnetic footpoint of Earth and the particle onsets for 26 eastern SEP events.

The other candidate to explain the release of particles is the expansion of the shock in the outer corona.
It is suggested that the particle release times are associated with the times when the shock establishes contact with the magnetic field lines connecting the spacecraft, rather than the arrival times of the EUV wave at the magnetic footpoints of the spacecraft.
\citet{Prise14} and \citet{Gomez-Herrero15} study the SEP event on 2011 November 3 with different methods.
They conclude that the EUV wave is too slow to explain the particle release, the expansion of the CME-driven shock at higher altitudes is consistent with the release times of particles at different spacecraft.
\citet{Kwon14} develop a geometrical model to determine the three-dimensional (3D) structure of the shock using the EUV and white-light (WL) observations from multipoint spacecraft.
\citet{Lario16} apply the geometrical model to fit the structure of the shock and indicate that the particles are released when the portion of the shock magnetically connected to each spacecraft is at a relatively high altitude.
Other physical mechanisms may also play a role in the particle release and the wide longitudinal extent of an SEP event.
\citet{Kouloumvakos16} discuss the role of the cross-field diffusion for the particle transport by tracking the EUV wave and modeling the shock on 2012 March 7.

In this paper, we study the release of energetic particles using coordinated imaging and in situ observations from \stereo{} and the near-Earth spacecraft.
The roles of the associated EUV wave and shock during the release of particles are still under debate.
It is necessary to investigate this problem using combined imaging and in situ data.
We focus on the 2012 January 27 solar eruption that originated from NOAA AR 11402 (N29\degree W86\degree).
It was associated with an X1.7 flare starting at 17:37 UT and peaking at 18:37 UT.
It produced an intense SEP event accompanied by a wide expanded EUV wave, halo CME and CME-driven shock.
The CME speed was about 2508 km s${{}^{-1}}$ estimated in the \soho/LASCO CME catalog (\url{https://cdaw.gsfc.nasa.gov/CME_list/}).
The center panel of Figure 1 shows the longitude of the active region (red arrow) and the positions of \stereoa, ${B}$ and Earth during this event.
\stereoa{} was located at 22\degree{} west of  the active region, \stereob{} was located at the opposite side of the Sun, and Earth was 86\degree{} east with respect to the active region.
These widely separated spacecraft offer EUV and WL observations for the associated EUV wave, CME and shock from three vantage points.
The active region was imaged by \stereoa{} on the solar disk, while the near-Earth spacecraft were in a good position to offer lateral imaging observations and measure the energetic particles (see Figure 1).
This configuration provides a good opportunity to track the evolution of the EUV wave in the corona and the shock at higher altitudes as well as to determine their connection with the particle release.
In Section 2, we describe the observations and perform the analysis.
The results are summarized and discussed in Section 3.

\section{Observations and Data Analysis}

\subsection{In Situ Measurements}

Figure 2 shows, from top to bottom, in situ solar wind speed, magnetic field magnitude and energetic proton intensities in different energy channels (three bottom panels) as measured by \stereob{} (left column), near-Earth spacecraft (middle column) and \stereoa{} (right column).
The solar wind data are measured by the Solar Wind Experiment \citep[SWE;][]{Ogilvie95} on board \wind{} and the Plasma and Suprathermal Ion Composition \citep[PLASTIC;][]{Galvin08} on board \stereo.
The magnetic field strength are obtained from the Magnetic Field Investigation (MFI) on \wind{} and the Magnetic Field Experiment \citep[][]{Galvin08} on board \stereo.
The near-Earth energetic particle intensities are observed by the Energetic Particle Sensors on board the \emph{Geostationary Operational Environmental Satellites} \citep[\goes{};][]{Sauer93}; the Energetic Relativistic Nuclei and Electron instrument \citep[ERNE;][] {Torsti95} on board \soho{}; and the Electron, Proton, and Alpha Monitor \citep[EPAM;][]{Gold98} on \ace.
The energetic particle intensities at \stereo{} are observed by the High-Energy Telescope \citep[HET;][]{Rosenvinge08}, Low-Energy Telescope \citep[LET;][] {Mewaldt08}, and the Solar Electron and Proton Telescope \citep[SEPT;][] {Muller08}.
Some of these instruments do not allow distinction between protons and ion species, therefore we will assume that most of the intensities plotted in Figure 2 are constituted by the most abundant protons and use interchangeably the term proton or particle in the following discussions.

The particle intensity-time profiles during the SEP event could provide information on the source of energetic particles and the features of the shock \citep[e.g.,][]{Reames96,Dresing12}.
The proton intensity profiles at \goes{} show a classic behavior of a western event as discussed by \citet{Cane88} and \citet{Reames99}.
Because of the active region being over the west limb as viewed from the near-Earth spacecraft, proton intensities of all energy channels at \goes{} rose and peaked quickly after the onset by around two orders of magnitude increase relative to the pre-event intensities.
Particle intensity at ${>}$100 MeV energies lasted for about 2 days above the pre-event intensity and decreased to the pre-event intensity level before the arrival of the shock.
The near-Earth spacecraft may be connected to the portion of the coronal shock with enough efficiency to accelerate particles to high energies directly after the solar eruption on January 27.
Particle intensity at ${>}$10 MeV energies showed a much longer decay even after the arrival of the shock (about 16:00 UT on January 30).
An interplanetary CME (ICME) interval was observed between 3:50 UT on January 31 and 23:50 UT on February 1 (gray shadow region in Figure 2).
A shock wave is indicated by sharp increases in the proton density, speed, temperature and magnetic field strength.
Signatures used to identify ICME from solar wind measurements mainly include depressed proton temperature, an enhanced helium/proton density ratio and smooth field rotation.
Note that the ${>}$10 MeV protons exhibited high intensity before the solar eruption.
The ${<}$1 MeV particle intensities at \ace{} rose slowly and peaked at the shock passage.
It indicates that the low energy particles were accelerated by the shock continuously.

As shown in the right panel of Figure 2, \stereoa{} observed a shock at about 13:13 UT on January 29, which was followed by an ICME between 7:20 UT on January 30 and 12:10 UT on February 2.
Significant increases of energetic particle intensities were observed by \stereoa /HET after the onset of the January 27 eruption.
Particle intensities of all energy channels observed by \stereoa{} peaked around the shock passage with two orders of magnitude increase relative to the pre-event intensity levels.
Note that the pre-event particle intensities at \stereoa{} were influenced by a prior event generated by a solar eruption on January 23 from the same active region NOAA 11402 at N29\degree W20\degree{} \citep{Liu13}.
The SEP event associated with the January 23 eruption was observed by \stereoa, ${B}$ and the near-Earth spacecraft and lasted until the onset of the January 27 solar event.

\stereob{} registered a weak shock on January 30 and observed slight increases of the energetic particle intensities about 12 hours after the shock arrival.
The portion of a shock without a persistent driver (i.e., CME) would decay quickly as it expands in the IP medium \citep{Liu09,Liu17,Kwon17}.
As the January 27 eruption is a backsided event for \stereob{} (see Figure 1), the shock would not be able to reach \stereob.
The imaging observations from \soho{} show that there was an eastward CME on January 26, which had a plane-of-sky speed of 431 km s${{}^{-1}}$ and was first seen at 19:36 UT in LASCO C2 (as listed in the \soho/LASCO CME catalog).
It was likely related to the shock and the increases of the energetic particle intensities observed by \stereob{}.
Consequently, only the near-Earth spacecraft and \stereoa{} observed the high-energy particles associated with the January 27 eruption.

\subsection{Calculation of Particle Release Times}

We apply two methods to determine the release times of the first arriving particles observed at L1 and \stereoa, the velocity dispersion analysis \citep[VDA; e.g.,][]{Huttunen-Heikinmaa05, Vainio13, Kouloumvakos15} and the time shifting analysis \citep[TSA; e.g.,][]{Vainio13, Kouloumvakos16,Lario17_Aug14}.
The VDA is a typical method to determine the particle release time and the apparent path length, whose performance has been evaluated in several studies \citep[e.g.,][]{Lintunen04,Saiz05,Rouillard12,Vainio13,Zhu16}.
The TSA method uses the onset time of the available highest energy channel of the particles and assumes that the first arriving particles observed by the spacecraft propagate scatter-freely along the nominal Parker spiral field lines.
The detail of the TSA algorithm has been discussed in \citet{Vainio13}.
A Poisson-CUSUM method \citep{Huttunen-Heikinmaa05} is used to determine the onset times of particles at different energies.
Note that the particle onset time identified by the Poisson-CUSUM method is sensitive to the mean value of the pre-event intensity.
Here we choose different values of the pre-event particle intensities for different energy channels in order to obtain relatively accurate onset times.

We display the result from the VDA method applied to the onset times of 1-minute resolution proton data from \soho/ERNE in Figure 3.
The method gives a release time of 18:25 UT ${\pm}$ 4 minutes (where 8 minutes have been added in order to compare with the imaging observations) and the corresponding path length of 1.5 ${\pm}$ 0.06 au.
We also obtain a release time of 18:38 UT ${\pm}$ 2 minutes (again, 8 minutes have been added) assuming a length of the nominal Parker spiral magnetic field line of 1.09 ${\pm}$ 0.01 au using the TSA method with the highest energy channel ranging from 101.0 to 131.0 MeV at \soho.
This release time is about 13 minutes later than the release time obtained from the VDA method.
A statistical analysis by \citet{Vainio13} shows that the VDA method provides reasonable release time relative to those from the TSA method in many cases.
Hence we use the release time of the particles observed at L1 determined with the VDA method.

For \stereoa, the pre-event intensities are 1${\sim}$2 orders of magnitude higher than the quiet period levels, except for the highest energy channel from 60.0 to 100.0 MeV.
The particle increases at the beginning of the SEP event are not obvious in several energy channels compared with the pre-event intensities.
Therefore, the VDA method could not be used to determine the release time of the particles at \stereoa{}.
We use the particle onset time of the highest energy channel (60-100 MeV) observed by \stereoa{} in the TSA method, and obtain a release time of 20:08 UT ${\pm}$ 5 minutes (8 minutes have been added) assuming a nominal length of the Parker spiral magnetic field line of 1.14 ${\pm}$ 0.01 au.
Using the results of 115 SEP events (we reject the SEP events whose particle release times between the VDA and TSA methods are beyond 2 hours) given by \citet{Vainio13}, we evaluate the mean delay between the VDA and TSA methods which is about 32 minutes.
Considering the effect of the prior event and the delay of the TSA result, we suggest a lower limit of 19:36 UT ${\pm}$ 5 minutes (i.e., subtract 32 minutes from the TSA result) for the release time of the particles observed at \stereoa.
It is much later than the release time of the particles at L1-observers (18:25 UT ${\pm}$ 4 minutes).
The delay of the particle release time of \stereoa{} may result from the fact that the magnetic connectivity between L1-observers and the portion of the shock with enough particle acceleration efficiency was established earlier.

\subsection{Magnetic Connectivity to the Sun}

Table 1 lists the locations of \stereoa, ${B}$ and Earth in the heliosphere and the footpoints of the magnetic field lines connecting the spacecraft to the Sun on January 27.
We also give the positions of STEREO and Earth in the center of Figure 1.
We estimate the magnetic footpoints of the spacecraft with different methods.
Columns 6-7 give the coordinates of the magnetic footpoints of the spacecraft estimated by tracing the nominal Parker spiral lines from the spacecraft to the solar surface.
We use the equation $\phi=\phi_{o}+\Omega L/V_{sw}$ for the estimation of the magnetic footpoint, where $\phi$ and $\phi_{o}$ are the longitudes of the magnetic footpoint and the spacecraft, respectively, $\Omega$ is the solar angular speed, ${L}$ can be considered as the radial distance from the spacecraft to the Sun, and $V_{sw}$ is the measured solar wind speed (as listed in column 5).
Columns 8-9 show the coordinates of the magnetic footpoints of the spacecraft estimated with a Potential Field Source Surface \citep[PFSS;][]{Schrijver03} method.
It includes two steps: first, we track the nominal Parker spiral line from the position of the spacecraft to a heliocentric distance of 2.5 R$_{\odot}$; then we estimate the footpoint by following the field line back to the solar surface based on the result of the PFSS model.

Figure 4 shows a \gong{} synoptic map at 17:44 UT on 2012 January 27 with open magnetic field lines simulated by the PFSS model and connected to the ecliptic plane (\url{http://gong.nso.edu/data/magmap/pfss.html}).
The locations of the spacecraft and their magnetic footpoints estimated with different methods are marked on the synoptic map.
A difference of about above 15\degree${ }$ in both longitude and latitude can be seen in the locations of the magnetic footpoints of L1-observers and \stereoa{} from the two methods.
Both methods provide similar longitudes for the magnetic footpoint of STEREO B.
All magnetic footpoints estimated by the PFSS method are located near the edges of coronal holes.
The result from the PFSS method is consistent with the suggestion that the release time of particles at the spacecraft is associated with the distance between the magnetic footpoint and the active region on the Sun \citep{Lario14}.
Note that the Parker spiral line method does not consider the complex magnetic configuration below 2.5 R$_{\odot}$ from the Sun.
In contrast, the PFSS method tracks the magnetic field line from 2.5 R$_{\odot}$ to the solar surface using the result of extrapolated field lines, which is based on the observed photospheric magnetic fields.
We apply the result estimated by the PFSS method to the analysis of the propagation of the EUV wave in the corona.

\subsection{EUV Wave Observations}

Figure 5 shows the observations of the EUV raw and running difference images from \sdo/AIA (211 \AA) and  the \emph{Extreme Ultraviolet Imager} \citep[EUVI;][] {Wuelser04} instrument on board \stereoa{} (195 \AA).
The magnetic footpoints of the L1-observers and \stereoa{}, estimated by the PFSS model, are marked by the yellow and blue crosses, respectively.
The white arrows indicate the EUV wave and the red arrows mark the active region associated with the January 27 solar event.
The solar eruption on January 27 is a western limb event in the field of view (FOV) of \sdo/AIA and a disk event for \stereoa/EUVI (see Figure 1).
Both views observed the EUV wave propagating from the active region.
The EUV wave propagated outward from the active region along almost all directions except the southern part of the disk as shown in \stereoa/EUVI and traveled to different directions as seen in \sdo/AIA.
Several studies suggest that the propagation of an EUV wave may be affected by active regions, coronal holes and other features it may encounter \citep[e.g.,][]{Thompson98, Ofman02, Gopalswamy09, Long11, Olmedo12}.
Note that two prominent coronal holes were observed by \sdo/AIA on the solar disk and some other active regions were located in the southward of the active region of interest as viewed from \stereoa/EUVI (see the left panels of Figure 5).
The coronal holes observed by \sdo/AIA may restrain the expansion of the EUV wave.
Simultaneously, the southern propagation of the EUV wave may be influenced by other active regions located on the southern side of the active region.
The EUV wave initially expanded consistently but was deformed since about 18:35 UT, and became diffuse with time.
There is no definite indication of EUV wave arrival at the magnetic footpoint of \stereoa{} or L1-observers.

In order to confirm whether the EUV wave has arrived at the magnetic footpoints of L1-observers and \stereoa, we track the motion of the EUV wave by stacking the EUV running difference images along the directions from the active region to the magnetic footpoints of the spacecraft, as shown by the black curves marked in the fourth column of Figure 5.
The stacking result of 211 \AA{} running difference images from \sdo{}/AIA and the result combing 195 \AA{} and 304 \AA{} running difference images from \stereoa{}/EUVI are shown in Figure 6.
The top edges of both panels indicate the magnetic footpoints of L1-observers (top panel) and \stereoa{} (bottom panel).
The bottom edges mark the locations near the active region.
The EUV wave is tracked by the red dashed curve in each panel.
Obviously, the track of the EUV wave front observed by \sdo{} stopped at about 18:35 UT.
The EUV wave propagation may be affected by the coronal hole around L1 magnetic footpoint.
The EUV wave did not arrive at the magnetic footpoint of L1-observers.
The track signature of the EUV wave observed by \stereoa{} became very faint before the particles observed by \stereoa{} were released (either using the lower limit of 19:36 UT ${\pm}$ 5 minutes or 20:08 UT ${\pm}$ 5 minutes release times). From the trace, we can see that the EUV wave, again, failed to arrive at the magnetic footpoint of \stereoa.

\subsection{Shock Observations and Modeling}

We determine the 3D structure of the shock by applying the geometrical ellipsoid model developed by \citet{Kwon14} to EUV and WL images from \sdo{}, \soho{} and \stereo.
The signature of the CME-driven shock in WL observations can be identified with a faint but relatively sharp brightness enhancement around the CME leading edge \citep[e.g.,][]{Vourlidas03, Liu08, Ontiveros09}.
The EUV wave is thought to track the lateral expansion of the coronal shock initially driven by the CME and propagate freely in the lower corona as the shock expands to higher altitude \citep[e.g.,][]{Patsourakos09,Kwon14, Miteva14, Kwon17}.
Figure 7 shows the geometrical fitting of the shock overplotted onto the running difference images by combing EUV and WL observations from three vantage points.
Seven free parameters are used in the geometrical ellipsoid model: the height, longitude, and latitude of the center of the ellipsoid, the lengths of the three semi-principal axes, and the rotation angle of the ellipsoid \citep{Kwon14}.
We adjust the free parameters and obtain a visually satisfied fit to match the images from all viewpoints simultaneously.

As shown in Figure 7, the footprint of the ellipsoid model in the solar surface coincides with the EUV wave front (see the top panel of \sdo{} and \stereoa).
The shock front is well represented by the ellipsoid model, expect for the portion along the westward direction of the Sun (as viewed from \stereob).
Note that the geometrical ellipsoid fitting is an idealized approximation for the shock.
The actual CME and shock could have complex shapes as they propagate and expand away from the Sun.
Nevertheless, the shock modeling provides key information for the connectivity between the shock and the spacecraft through magnetic field lines.
We start the shock fitting at 18:15 UT and obtain a number of fitting results until 20:06 UT.
The fitting becomes difficult as the signature of the shock becomes diffuse in the coronagraph images later.
Figure 8 shows the radial distance and the speed of the shock nose.
The speed of the shock nose was first accelerated to ${\sim}$3300 km s${{}^{-1}}$ at around 18:40 UT, when the shock nose reached a heliocentric height of ${\sim}$6 R$_{\odot}$.
The average speed is about 2300 km s${{}^{-1}}$, comparable to the CME speed of about 2508 km s${{}^{-1}}$ in the LASCO CME catalog.
The speed showed a deceleration till ${\sim}$18:55 UT.
The directional axis of the fitted shock turned from N29\degree W79\degree{} at ${\sim}$18:15 UT to N23\degree W55\degree{} at ${\sim}$19:24 UT, and kept this direction as the shock propagated outward from the Sun at least until 20:06 UT.

We present the contact between the fitted shock and the magnetic field lines connecting L1-observers and \stereoa{} from different views in Figure 9.
The point of intersection between the shock front and the field line connecting to a given observer has been named ``cobpoint" \citep[i.e., Connecting with the OBserver point;][]{Heras95}.
The views in the top row are seen from L1-observers (left panel) and \stereoa{} (middle and right panels) and those in the bottom row are projections in the plane determined by the directional axis of the ellipsoid and the cobpoint of the spacecraft.
The magnetic field lines connecting L1-observers and \stereoa{} are denoted by the black and red lines, respectively.
The magnetic field lines are obtained from the PFSS method as discussed in Section 2.3.
Black arrows in the bottom row indicate the normal of the shock at the cobpoints.
The technique of the shock fitting has also been applied to study the SEP events on 2014 February 25 and 2010 August 14 \citep{Lario16,Lario17_Aug14}.
Note that the portion of the shock along the westward direction as seen from \stereob{} (left panel of Figure 7) is not fitted well.
The magnetic contact between the fitted shock and \stereob{} could not indicate the actual situation.
So we just discuss the magnetic contact of L1-observers and \stereoa{} here.
The height of the cobpoint  and the orientation of the magnetic field with respect to the normal to the shock surface ($\theta_\mathrm{Bn}$) could be identified from the geometrical fitting of the shock.
Furthermore, it allows us to evaluate the Mach number of the shock at the cobpoints when the particles were released.
We compute the shock speed ($V_\mathrm{s}$) using the three-point Lagrangian interpolation to fit the heights of the discrete cobpoints with time along the normal direction of the shock. The fast mode speed is defined as:
\begin{equation}
    V_\mathrm{FM} = \sqrt{ \frac{1}{2} [ V_\mathrm{A}^2 + C_\mathrm{s}^2 + \sqrt{ (V_\mathrm{A}^2 + C_\mathrm{s}^2)^2 - 4 V_\mathrm{A}^2 C_\mathrm{s}^2 cos^2 \theta_\mathrm{Bn} } ] },
\end{equation}
where $V_\mathrm{A}$ is the Alfv\'{e}n speed ($V_\mathrm{A} = B/ \sqrt{4 \pi \rho}$), $C_\mathrm{s}$ is the sound speed of ${\sim}$180 km s${{}^{-1}}$ for a coronal temperature of 1.4 MK.
In particular, the magnetic field strength (${B}$) and the mass density ($\rho$) are derived by the empirical models from \citet{Mann99} and \citet{Leblanc98}, respectively.
Compared with the coronal density models from \citet{Newkirk61} and \citet{Saito77}, the \citet{Leblanc98} model provides a lower electron density.
Here we give a crude lower limit of the shock Mach number, i.e., $M_\mathrm{FM} = V_\mathrm{s} / V_\mathrm{FM}$, for the cobpoints at the particle release times.

As shown in the left panel of Figure 9, the magnetic field line of L1-observers first connected to the flank of the shock when the particles at L1-observers were released (18:25 UT ${\pm}$ 4 minutes).
It means that the particles observed at L1 were released as L1-observers became connected with the coronal shock via the magnetic filed line.
The fitting result gives a cobpoint height of 2.4${\pm}$0.5 R$_{\odot}$ from the heliocenter, where the shock is oblique ($\theta_{Bn}$ $\sim$51${\pm}$5\degree) and the estimated $M_\mathrm{FM}$ is ${\sim}$1.5${\pm}$0.5 at the particle release time of L1-observers.
The uncertainties of these parameters correspond to their changes during the time interval given by the error of the particle release time.
\stereoa{} was first connected to the backside of the shock at $\sim$18:46 UT.
This is earlier than the estimated release time of the particles observed at \stereoa.
The Mach number for the initial contact of \stereoa{} is generally below 1.
The cobpoint of \stereoa{} is at a heliocentric height of 8.2${\pm}$0.7 R$_{\odot}$, where $\theta_{Bn}$ is 33${\pm}$1\degree{} and the $M_\mathrm{FM}$ is ${\sim}$1.7${\pm}$0.2 at the lower limit of the particle release time of \stereoa{} (19:36 UT ${\pm}$ 5 minutes).
Subsequently, a heliocentric height of ${\sim}$11.3 R$_{\odot}$, $\theta_{Bn}$ of ${\sim}$31\degree{} and $M_\mathrm{FM}$ of ${\sim}$1.5 are derived for the cobpoint of \stereoa{} at the particle release time based on the TSA method (20:08 UT).
A possible explanation for the delay of the particle release at \stereoa{} is that at the initial contact the part connected to \stereoa{} was not really a shock, but perhaps just a wave expanding backward toward the Sun \citep{Liu17}.
Particles could be released when \stereoa{} was connected to the portion of the shock with enough particle acceleration efficiency.
The Mach number is often used as a key parameter to evaluate the particle acceleration efficiency of the shock.
It reached a modest value of ${\sim}$1.5 at the cobpoints of both L1-observers and \stereoa{} when the particles were released.
Additionally, we should keep in mind that the magnetic field lines in Figure 9 are an approximate estimate obtained from the solar wind speed and the magnetic configuration before the solar eruption.
The actual field configuration could be complicated as the shock expanded away from the Sun.
Similarly, the empirical models used to estimate the Alfv\'{e}n speed and sound speed in Equation (1) are approximations that may not be representative of the state of the actual corona when this event occurred.

\section{DISCUSSION AND CONCLUSIONS}

The solar eruption on 2012 January 27 was associated with an X1.7 flare, EUV wave, CME-driven shock and wide-spread SEP event.
Combining multipoint remote sensing and in situ observations, we have estimated the release times of the particles observed by the spacecraft and the locations of the magnetic footpoints of the spacecraft, tracked the evolution of the EUV wave along the directions from the active region to the magnetic footpoints of the spacecraft, modeled the shock and identified the magnetic connectivities between the shock and the spacecraft at the estimated particle release times.
The results are summarized and discussed as follows.

Near-Earth spacecraft and \stereoa{} observed the shock in situ and the SEP event associated with the January 27 solar event, which is a backsided event for \stereob{}.
The high-energy particle intensities at both L1-observers and \stereoa{} exhibited almost immediate onsets, obvious increases and long durations above the pre-event intensity levels.
The particle intensities at \goes{} peaked quickly after the onset and those at \stereoa{} peaked around the shock passage with two orders of magnitude higher than the pre-event intensity levels.
Slight enhancements of energetic particle intensities at  \stereob{} were associated with another unrelated solar event.
The release time of the particles at L1-observers determined with the VDA method corresponded to the time when the magnetic field line of L1-observers initially connected to the shock.
We suggest that the L1-observers was directly connected to the portion of the shock with enough particle acceleration efficiency via the magnetic field line.
We give a lower limit of the particle release time for \stereoa{} based on the TSA method.
The particle release of \stereoa{} was later than the release of the particles at L1-observers.
It may be due to the fact that the magnetic connectivity between \stereoa{} and the portion of the shock with enough particle acceleration efficiency occurred later.

Determining the arrival of EUV waves at the magnetic footpoints of the spacecraft is essential to test whether the EUV wave is responsible for the particle release.
The EUV observations show that coronal holes and active regions influenced the propagation of the EUV wave, and no definite EUV wave arrived at the estimated magnetic footpoints of the spacecraft.
We track the evolution of the EUV wave along the paths from the active region to the magnetic footpoints of L1-observers and \stereoa.
It shows that the propagation of the EUV wave toward the magnetic footpoint of L1-observers was halted at about 18:35 UT.
The EUV wave may be restrained by the coronal hole around the magnetic footpoint of L1-observers.
The signature of the EUV wave toward the magnetic footpoint of \stereoa{} became diffuse and disappeared before the release of the particles at \stereoa.
As a result, we do not obtain any definite evidence in the current case to support the argument that the particle release is accounted for by the arrival of the EUV wave at the magnetic footpoint of the spacecraft.
This is consistent with previous studies of \citet{Gomez-Herrero15} and \citet{Lario14, Lario16}.
However, we cannot exclude the possibility that EUV waves contribute to the particle releases in some cases.
They may be able to explain the release of the first particles in those cases, but not the continuous releases which should be produced by the connection between the shock with enough particle acceleration efficiency and the observers.

The geometrical modeling of the shock gives the 3D shock structure and allows us to analyze its contact with the magnetic field lines of the spacecraft.
\citet{Lario16} suggest a relatively high altitude (${\gtrsim}$2 R$_{\odot}$ above the solar surface) for the height of the cobpoint when the particle were released.
The fitting result in the current case gives relative lower and higher altitudes, i.e., a heliocentric height of 2.4${\pm}$0.5 R$_{\odot}$ for the L1 cobpoint when the particles were released and of 8.2${\pm}$0.7 R$_{\odot}$ for \stereoa{} cobpoint at the lower limit of the particle release time.
The time when the magnetic field line of L1-observers initially connected to the shock was in good agreement with the release time of the particles observed at L1.
In contrast, the release time of the particles at \stereoa{} was later than the time when the magnetic connectivity between \stereoa{} and the shock was first established.
We suggest that \stereoa{} was first connected to a wave propagating backward toward the Sun via the magnetic field line, not the part of the shock with enough particle acceleration efficiency \citep{Liu17}.

Particle acceleration efficiency of the shock is correlated with the parameters including the density compression ratio, the angle between the shock normal and the upstream magnetic field, the Mach number and the position angle relative to the shock leading edge \citep[e.g.,][]{Bemporad11, Kwon17, Kwon18, Lario17_Mach}.
The fitting result shows $\theta_{Bn}$ of $\sim$51\degree{} and $\sim$31\degree{} at the cobpoints of L1-observers and \stereoa{} when the particles were released.
The Mach number evaluated by the empirical models gives a modest value of ${\sim}$1.5 for the cobpoints of both L1-observers and \stereoa{} at the particle release times.
It is comparable to the critical Mach number provided by \citet{Edmiston84}.
We suggest that the particle acceleration efficiency of the portion of the shock connected to the spacecraft determines the release of energetic particles at the spacecraft.
We confirm that the propagation of the EUV wave is affected by coronal holes and other active regions and it is not responsible for the particle release in our case.
A possible role of particle perpendicular diffusion may contribute to the release of particles observed at the spacecraft, but this is beyond the scope of this paper.

\acknowledgments The research was supported by NSFC under grants 41774179 and  41374173, and the Specialized Research Fund for State Key Laboratories of China. We acknowledge the use of the data from \stereo, \gong, \goes, \wind, \ace{} and \soho. We also thank Huidong HU, Wen HE and Xiaowei ZHAO for their helpful discussions in completing this paper.

 \newpage
 \bibliography{references}
 \bibliographystyle{apj}

\clearpage

\begin{table}
\caption{Spacecraft and their Magnetic Footpoint Locations. \label{propa}}
\begin{tabularx}{\linewidth}{@{\extracolsep{\fill}}lccccccccc}
\hline\hline
   & & & & &\multicolumn5c{Magnetic Footpoint Coordinates}\\
\cline{6-10}
  & \multicolumn{3 }{c}{S/C Location\tablenotemark{a}} & ${Vsw\tablenotemark{b}}$ & \multicolumn{2 }{c}{Parker Spiral\tablenotemark{c} }& & \multicolumn{2 }{c}{PFSS\tablenotemark{d}}\\
  \cline{2-4}
  \cline{6-7}
  \cline{9-10}
  \multicolumn1c{Spacecraft} & \multicolumn1c{R(AU)}  & \multicolumn1c{Lon} & \multicolumn1c{Lat} &  \multicolumn1c{($\rm km~s^{-1}$)} & \multicolumn1c{Lon} & \multicolumn1c{Lat} &   &\multicolumn1c{Lon} & \multicolumn1c{Lat}\\
  \multicolumn1c{(1)} &\multicolumn1c{(2)}&\multicolumn1c{(3)}&\multicolumn1c{(4)} &\multicolumn1c{(5)} &
  \multicolumn1c{(6)}&\multicolumn1c{(7)}&&\multicolumn1c{(8)} &\multicolumn1c{(9)}\\
  \hline
  \stereob{} & 1.06 &  -114 &    7 &  452&  -57 &  7 & &  -56 &  17   \\
          Earth & 0.98 &       0 &   -6 &  532&   45 &  -6 & &   31 &   -21   \\
  \stereoa{} & 0.96 &  108  &   -3 &  422&  164 &  -3 & &   -175 &   22   \\
\hline
\end{tabularx}
\tablenotetext{a}{The heliocentric radial distance (R), the heliographic inertial longitude (Lon) and the heliographic inertial latitude (Lat) of \stereoa, ${B}$ and Earth. }
\tablenotetext{b}{The average solar wind speeds V$_{sw}$ measured at the onset of the SEP event. }
\tablenotetext{c}{The coordinates of magnetic footpoints estimated with the Parker spiral field line method. }
\tablenotetext{d}{The coordinates of magnetic footpoints estimated with the PFSS method. }
\end{table}
\clearpage

\begin{figure}[!htb]
\centering
\noindent\includegraphics[width=40pc]{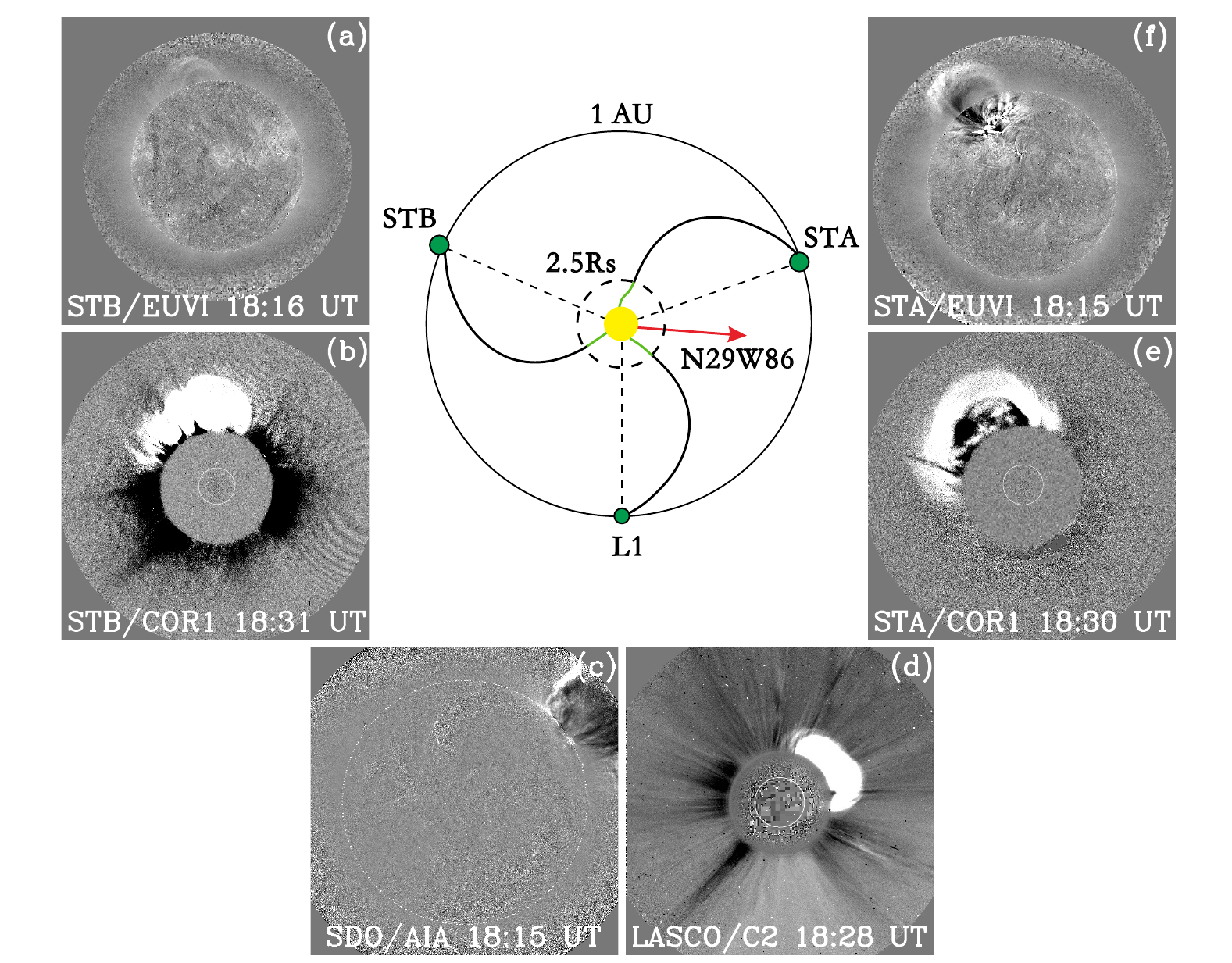}
\caption{
Positions of the spacecraft and associated observations on 2012 January 27.
The red arrow marks the longitude of the active region.
The black circle indicates the orbit of Earth, and the dash black circle shows the heliocentric distance of 2.5 R$_{\odot}$ (not to proportion).
The spiral lines represent the sketches of the magnetic field lines connecting the spacecraft with the Sun.
(a, c, f): running difference images of the solar eruption from \stereob/EUVI (195 \AA), \sdo/AIA (211 \AA) and \stereoa/EUVI (195 \AA), respectively.
(b, d, e): running difference images of the CME from \stereob/COR1, \soho/LASCO C2 and \stereoa/COR1, respectively.}
\end{figure}
\clearpage

\begin{figure}[!htb]
\centering
\noindent\includegraphics[width=40pc]{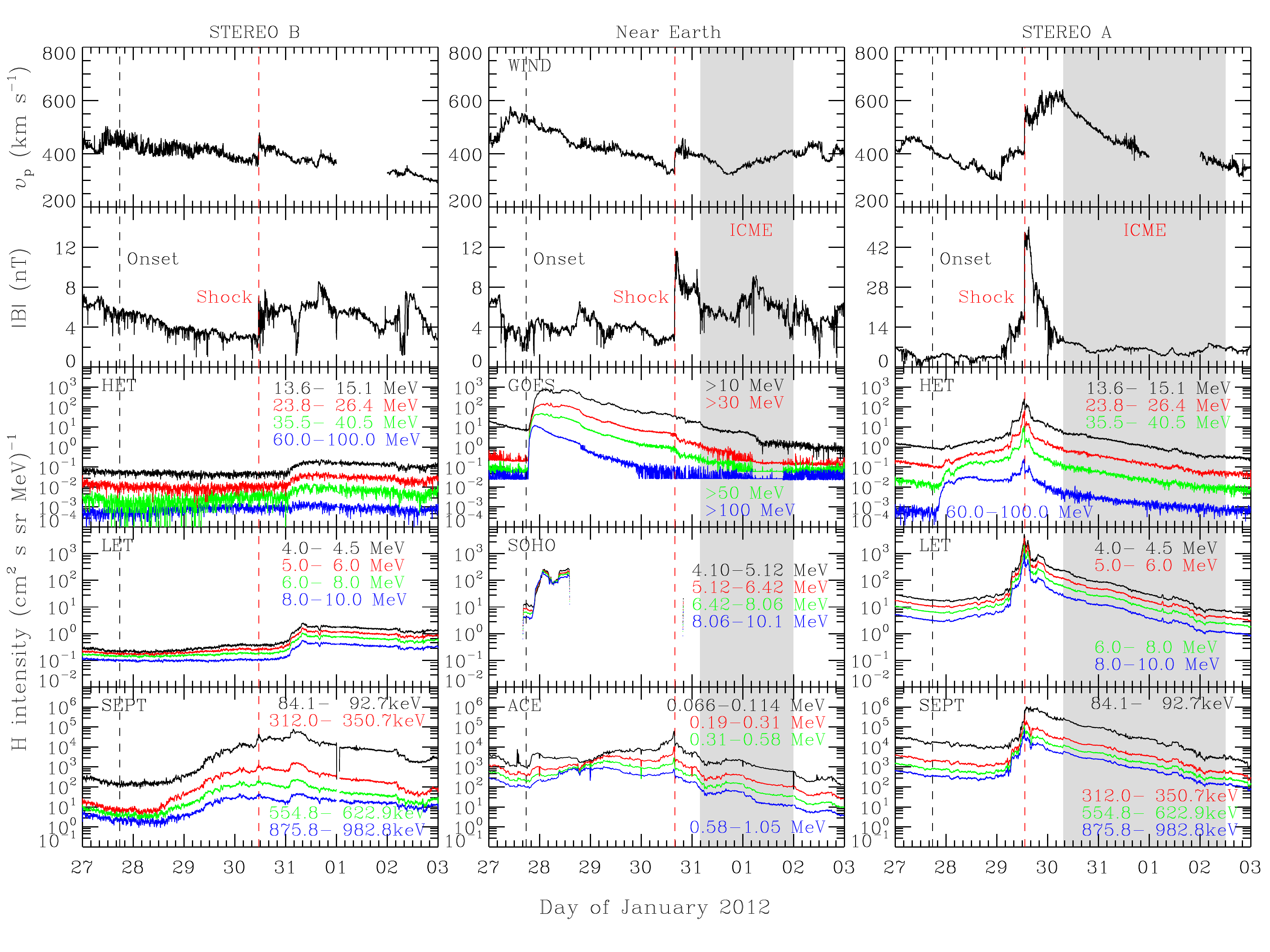}
\caption{Solar wind speed, magnetic field and proton intensities in different energy channels measured by \stereob{} (left), near-Earth spacecraft (middle) and \stereoa{} (right).
Dashed lines mark the time of the solar eruption (black) and the arrival times of the shocks (red), respectively.
The shaded region indicates the interval of the ICME.
The ICME was not observed at \stereob.}
\end{figure}
\clearpage

\begin{figure}[!htb]
\centering
\noindent\includegraphics[width=25pc]{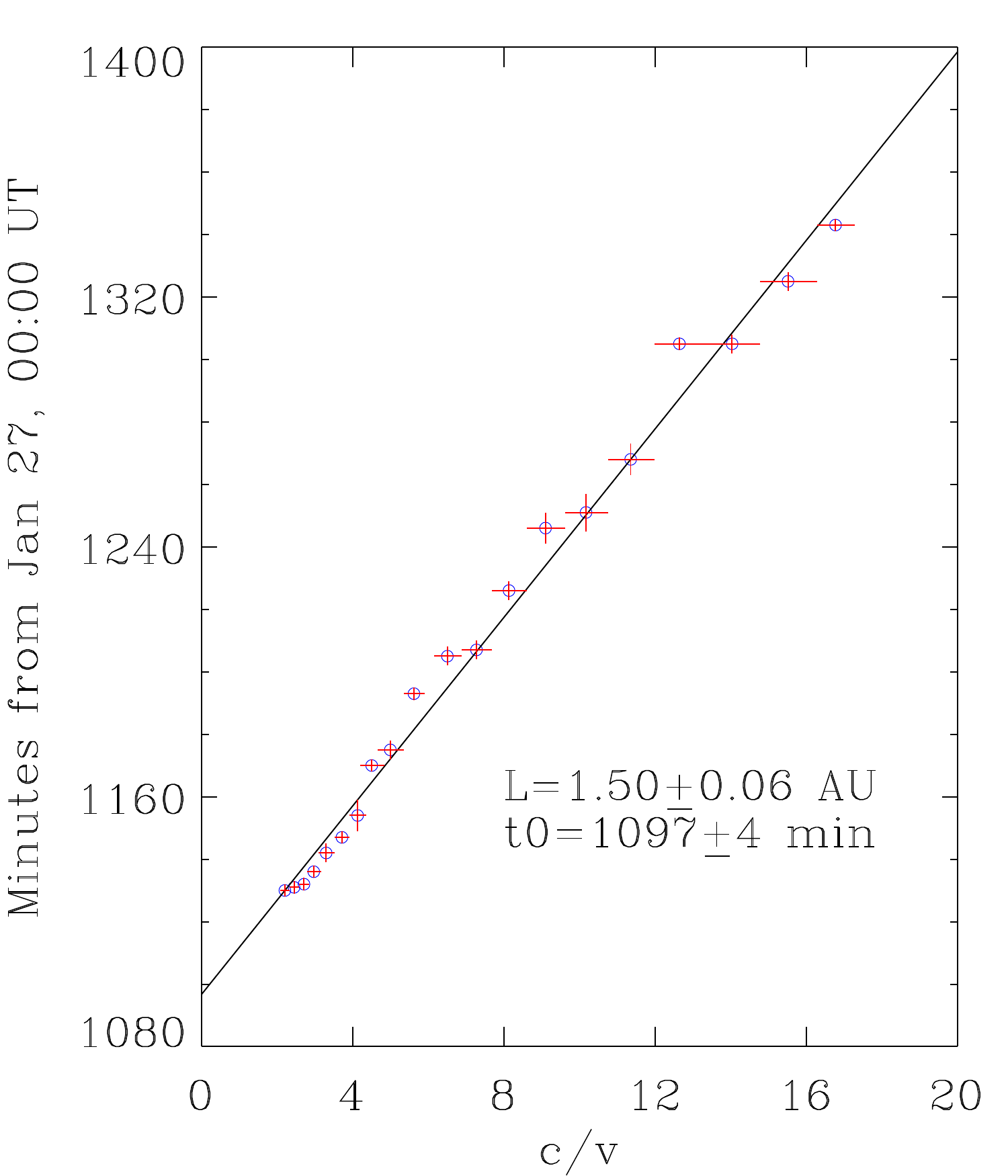}
\caption{Velocity dispersion analysis for \soho{} data.
The proton onset times observed by ERNE are shown as blue circles with red error bars.
The black line is a linear fit to all points.
The particle release time near the Sun with respect to the beginning of January 27 and the path length from the fitting are also given in the panel.}
\end{figure}
\clearpage

\begin{figure}[!htb]
\centering
\noindent\includegraphics[width=40pc]{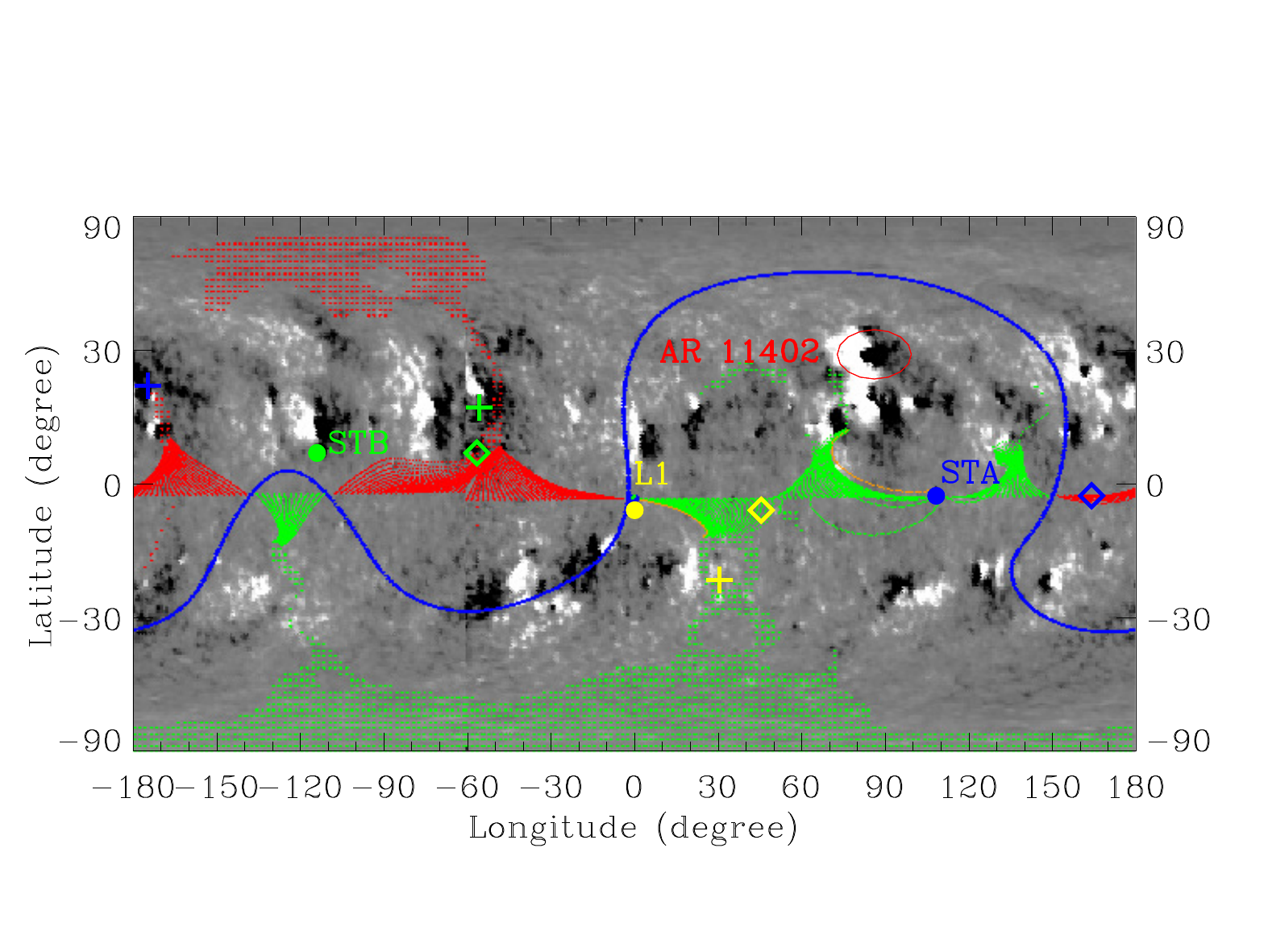}
\caption{
GONG synoptic map with the PFSS open magnetic field lines connecting the ecliptic plane on 2012 January 27.
Red dots denote the negative polarity and green dots mark the positive polarity.
The blue solid line is the heliospheric current sheet.
The green, yellow and blue filled circles indicate the projections of \stereob, L1-observers and \stereoa.
The magnetic footpoints of the spacecraft on the solar surface estimated by tracing the nominal Parker spiral lines are marked by the corresponding colored diamonds, and the magnetic footpoints estimated using the result of the PFSS model are presented by the corresponding colored crosses.
The active region is marked by a red ellipsoid.}
\end{figure}
\clearpage

\begin{figure}[!htb]
\centering
\noindent\includegraphics[width=40pc]{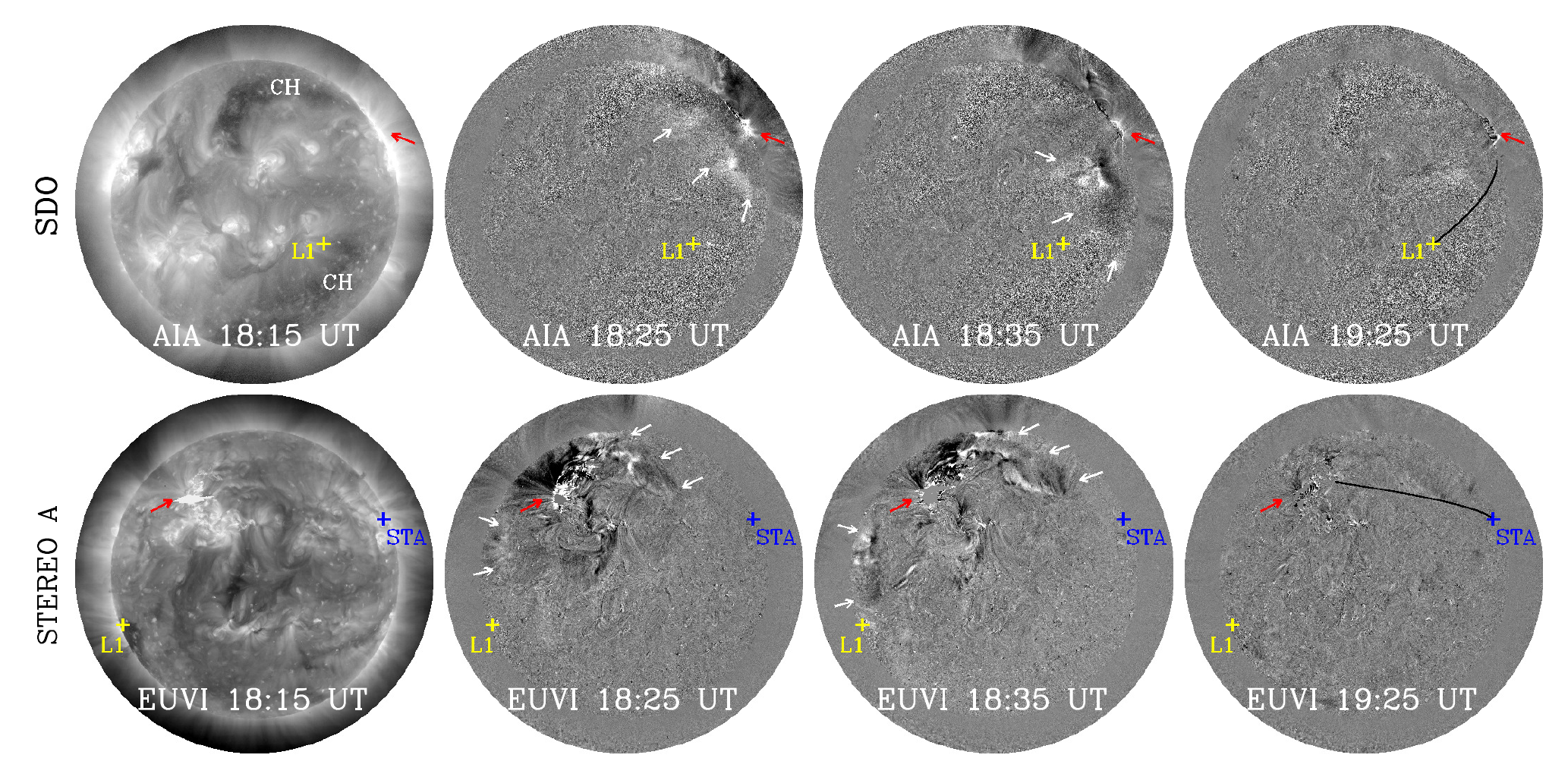}
\caption{Raw and running difference images observed by \sdo{}/AIA at 211 \AA{} (top) and \stereoa{}/EUVI at 195 \AA{} (bottom).
The coronal holes (CHs) are marked in \sdo{} raw image.
The red arrows mark the active region of interest and the white arrows indicate the EUV wave.
The yellow and blue crosses represent the magnetic footpoints of L1-observers and \stereoa{}, respectively.
 The black curves in the fourth column mark the traces chosen for the stacking slit of the EUV wave.
}
\end{figure}
\clearpage

\begin{figure}[!htb]
\centering
\noindent\includegraphics[width=30pc]{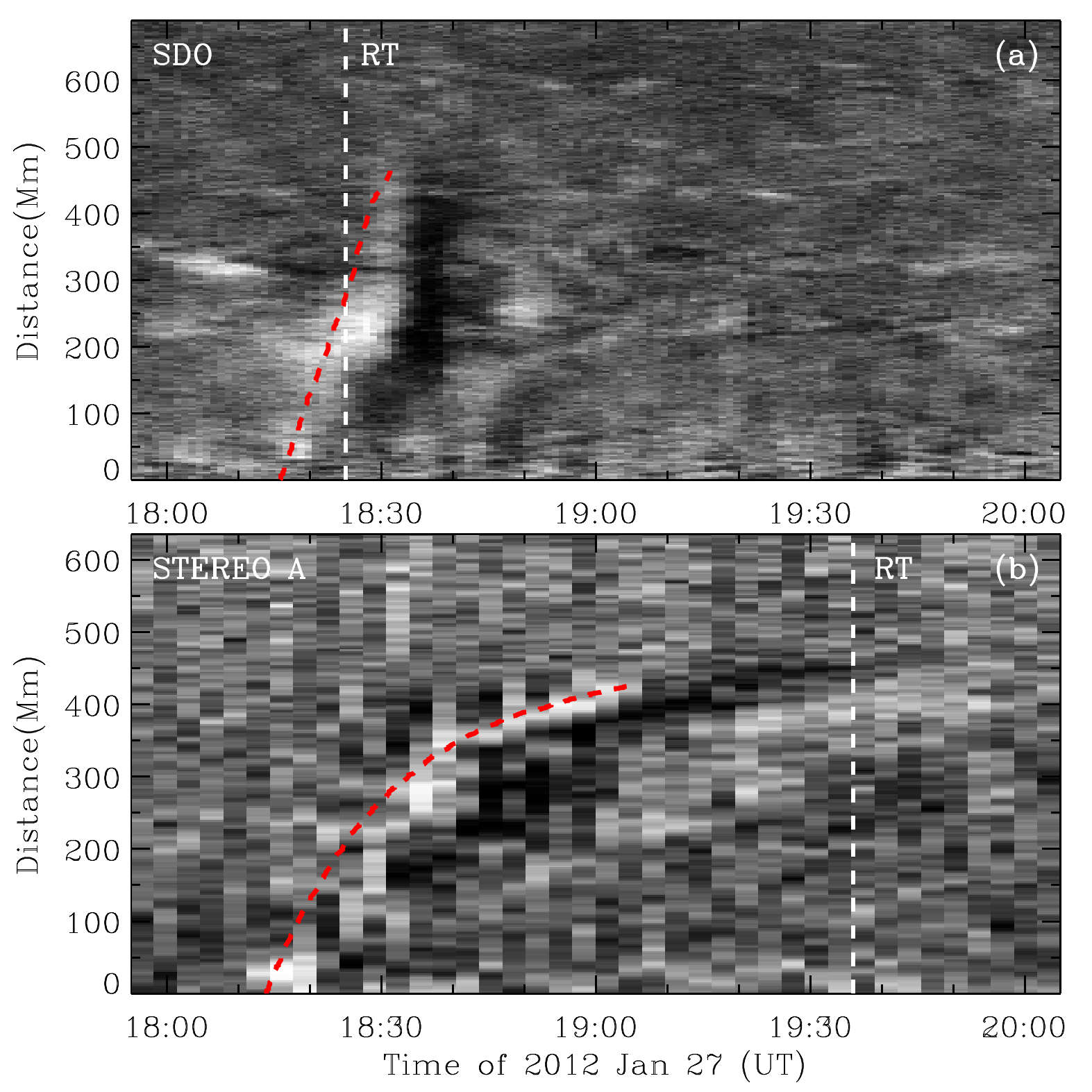}
\caption{Distance-Time plots by stacking the EUV running difference images (\sdo{}/AIA at 211 \AA{}, \stereoa{}/EUVI at 195 \AA{} and 304 \AA{}) along the tracks marked in the fourth column of Figure 5.
The bottom edges of both panels are the slit ends near the active region.
The top edges are the magnetic footpoints of L1-observers (top) and \stereoa{} (bottom).
The white vertical dashed lines indicate the estimated release time of the particles observed by the near-Earth spacecraft (18:25 UT) and a lower limit time of the particle release at \stereoa{} (19:36 UT), respectively.
}
\end{figure}
\clearpage

\begin{figure}[!htb]
\centering
\includegraphics[width=40pc]{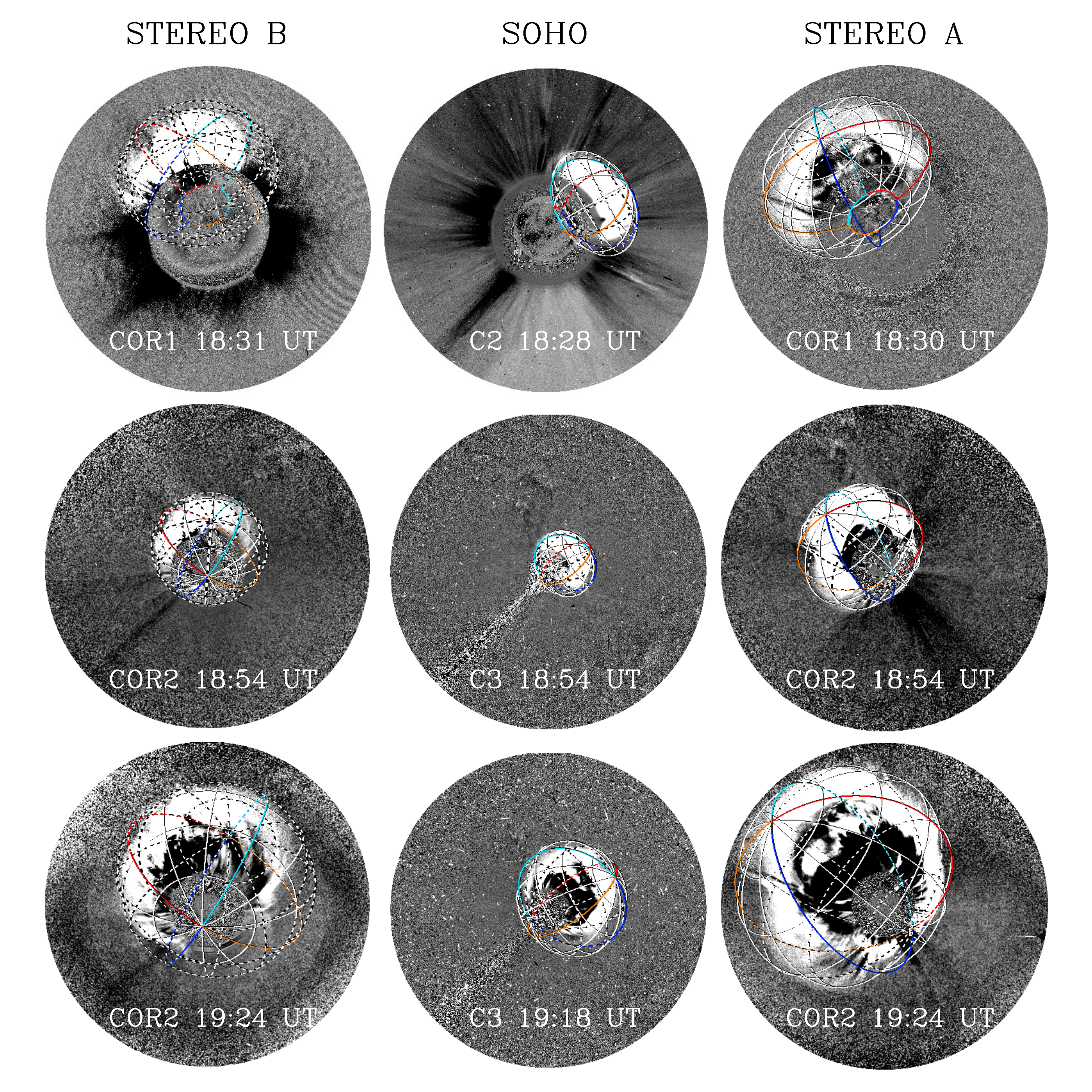}
    \caption{Running difference images on 2012 January 27 viewed from \stereob{} (left), \soho{} and \sdo{} (middle) and \stereoa{} (right) at different times.
The ellipsoid frame in each image shows the geometrical modeling of the shock front with the ellipsoid model developed by \citet{Kwon14}.
The colored lines (red, orange, blue and cyan) are used to represent different quadrants of the ellipsoid, and the white lines construct the surface of the ellipsoid (dashed lines mark the structure on the other side of the image plane).}
\end{figure}
\clearpage

\begin{figure}[!htb]
\centering
\noindent\includegraphics[width=38pc]{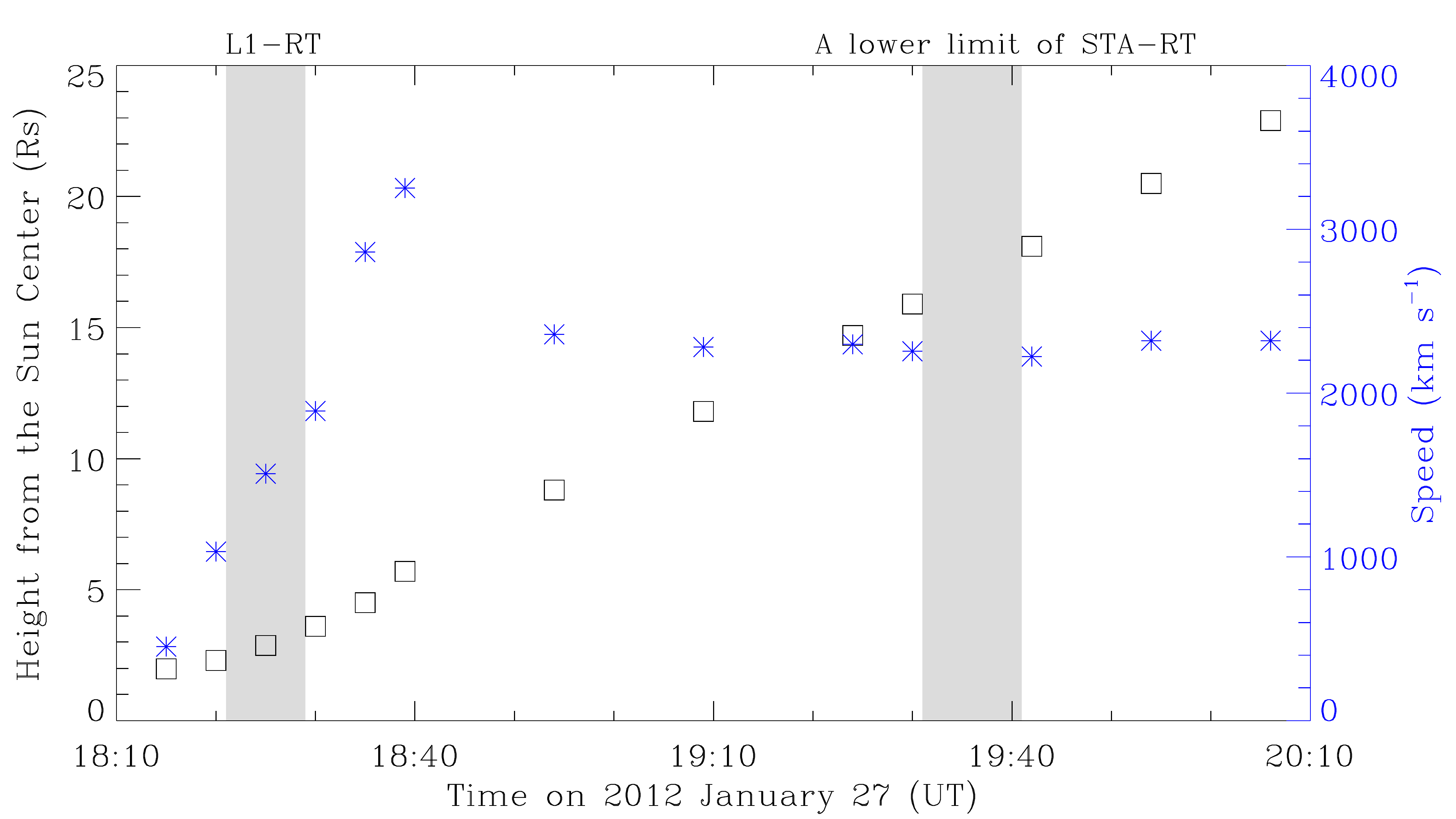}
\caption{Radial height and speed of the shock nose.
The speeds are calculated from adjacent distances using a numerical differentiation method with three-point Lagrangian interpolation.
The shaded regions indicate the estimated release time of the particles observed at L1 and a lower limit of the particle release time of \stereoa, respectively.}
\end{figure}
\clearpage

\begin{figure}[!htb]
\centering
\noindent\includegraphics[width=38pc]{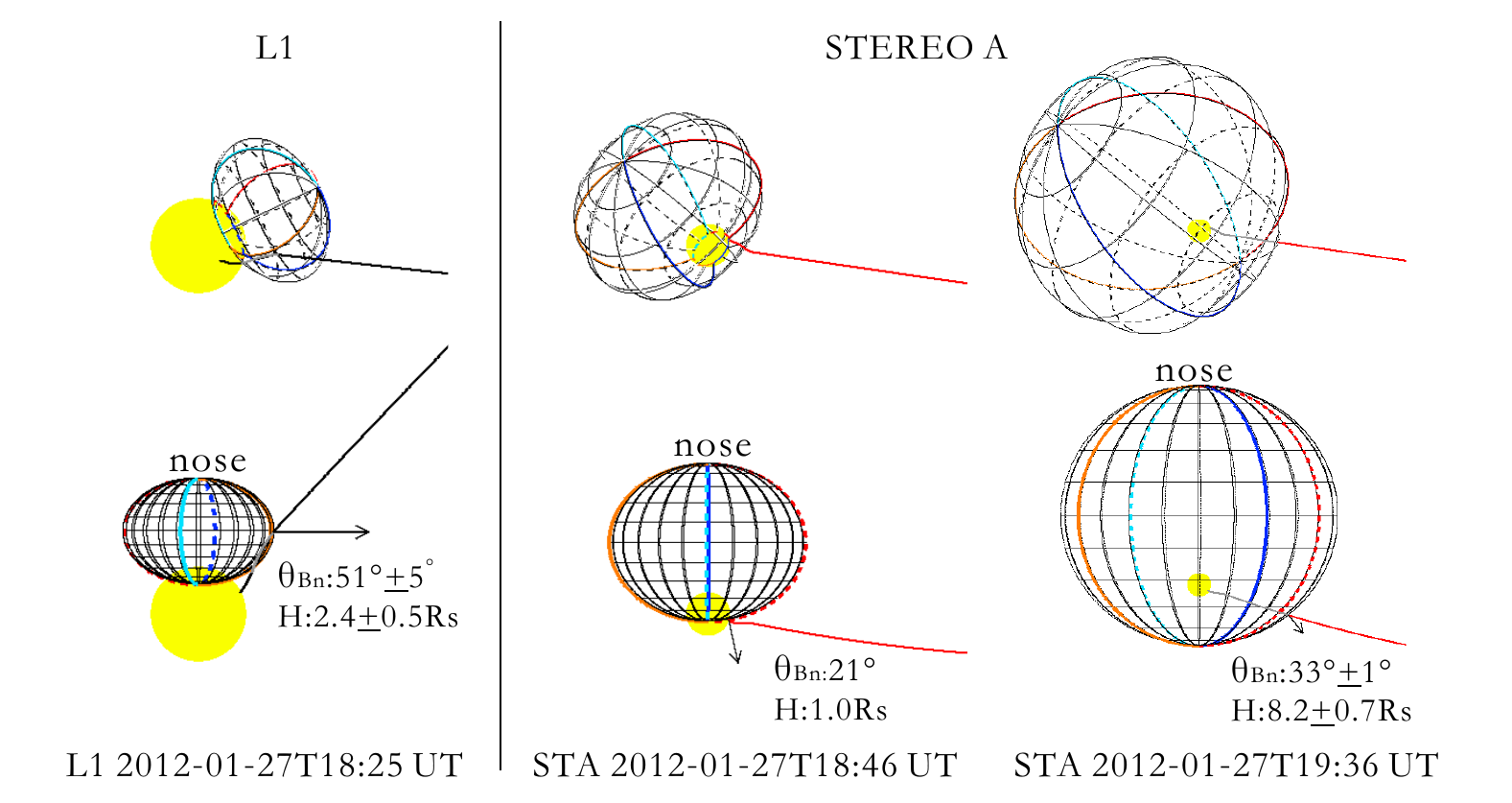}
\caption{
The ellipsoid shock as seen from the observers (L1 and \stereoa; top row) and projections in the plane formed by the Connecting-with-the-OBserver point (cobpoint) and the directional axis of the fitted shock (bottom row).
The yellow filled circle is the solar disk.
The black and red lines indicate the magnetic field lines connecting L1 and STEREO A from different views, respectively.
The arrows represent the normal of the shock at the cobpoints.
The angles between the shock normal and the magnetic field and the heights of the cobpoints are also given in the bottom row.
The gray lines connecting the solar disk are used for the portion of the field lines inside the fitted ellipsoid.}
\end{figure}
\clearpage

\end{document}